# Liquid Metal Enabled Droplet Circuits

Yi Ren [1] and Jing Liu [1, 2, 3*]

[1] Department of Biomedical Engineering & Future Lab, Tsinghua University, Beijing 100084, China;

[2] Technical Institute of Physics and Chemistry, Chinese Academy of Sciences, Beijing 100190, China;

[3] School of Future Technology, University of Chinese Academy of Sciences, Beijing 100049, China.

*Corresponding author. Email: jliubme@tsinghua.edu.cn.

**Abstract**

Conventional electrical circuits are generally rigid in their components and working styles which are not flexible and stretchable. From an alternative, liquid metal based soft electronics is offering important opportunities for innovating modern bioelectronics and electrical engineering. However, its running in wet environments such as aqueous solution, biological tissues or allied subjects still encounters many technical challenges. Here, we proposed a new conceptual electrical circuit, termed as droplet circuits, to fulfill the special needs as raised in the above mentioned areas. Such unconventional circuits are immersed in solution and composed of liquid metal droplets, conductive ions or wires such as carbon nanotubes. With specifically designed topological or directional structures/patterns, the liquid metal droplets composing the circuit can be discretely existing and disconnected from each other, while achieving the function of electron transport through conductive routes or quantum tunneling effect. The conductive wires serve as the electron transfer stations when the distance between two separate liquid metal droplets is far beyond than that quantum tunneling effects can support. The unique advantage of the current droplet circuit lies in that it allows parallel electron transport, high flexibility, self-healing, regulativity and multi-point connectivity, without needing to worry about circuit break. This would extend the category of classical electrical circuits into the newly emerging areas like realizing room temperature quantum computing, making brain-like intelligence or nerve-machine interface electronics etc. The mechanisms and potential scientific issues of the droplet circuits are interpreted. Future prospects along this direction are outlined.

**Keyword:**   Droplet circuits; Liquid metal; Quantum tunneling effect; Solution electronics; Electron transport; Ionic conduction; Quantum computing; Brain-like intelligence.

## 1. Introduction

Since the origination of electricity, it has become a necessity in daily life. Generally speaking, electrical circuit is rigid and continuous in its structures and components. Print circuit board (PCB) has been commonly used in various situations. However, classical rigid circuits could not easily adapt to human body due to poor flexibility and biocompatibility, limiting its value in biomedical and health care field. The increasing advancement of wearable devices and implantable systems leads to significant growth in flexible electronics [1]. Polymer nanomaterials, silk fibroin and liquid metal are being gradually adopted in soft electronics. Another main development trend in artificial circuits is molecular electronics. It was first proposed in 1974 by Arieh Aviram and Mark Ratner [2], which refers to a field that seeks to fabricate electrical devices and circuits with single molecules and molecular monolayers [3-6]. The fabrication of molecular electrical devices includes single-molecule break junctions and molecular monolayer devices [3]. Tailored by chemical design and synthesis, the function of molecular components can be rather diverse. Up to now, molecular components including diodes (Fig. 1(C)), switches, memory and transistors have been intensively researched [5]. Those components can be combined to construct molecular-scale electronic computers [7]. Based on the electrical properties of the molecular diode switches, quantum mechanical calculation can also possibly be performed.

Besides those artificial circuits, electrical circuits in fact intrinsically exist throughout human body. For instance, Dejean et al recently studied the neural circuits and cell types that mediate conditioned fear expression and recovery [8]. Moreover, voltage-gated channels are important switches for signal transportation in central and peripheral neural systems [9]. Bearing the above facts in mind and without losing generality, we can divide electrical circuits into two main categories: biologically inspired natural circuits and human made artificial circuits. One more trend as indicated in Fig. 1(E) is that, a new category is emerging to combine the naturally occurring neural circuits and the artificial circuits together to carry out complex functions. The core of such circuits can generally be called as brain-computer interfaces (BCIs) [10-13]. Clearly, BCIs are significant for those patients with serious disabilities such as tetraplegia and stroke. They especially mean a lot for future human needs in extending the limit of biological capability. At this stage, rapid serial visual presentation based BCIs have already been used to detect and recognize objects, providing a viable approach to prompt human-machine systems [13].

Despite the widespread application of currently available circuits, they all unavoidably encounter the possibility of circuit break, which would severely affect the normal operation of devices. As a remedy, we are dedicated here to propose an unconventional concept of electrical circuit to tackle the above challenges which can be named as droplet circuits. Such electrical circuit is enabled from the liquid metal droplets (LMDs) and is conductive in discontinuous form. The structure and working style of this kind of circuit are highly analogy and similar to that of the neuro network system. Therefore it is expected to be very useful in innovating the newly emerging areas like room temperature quantum computing, brain-like intelligence, and bran-machine interface etc.

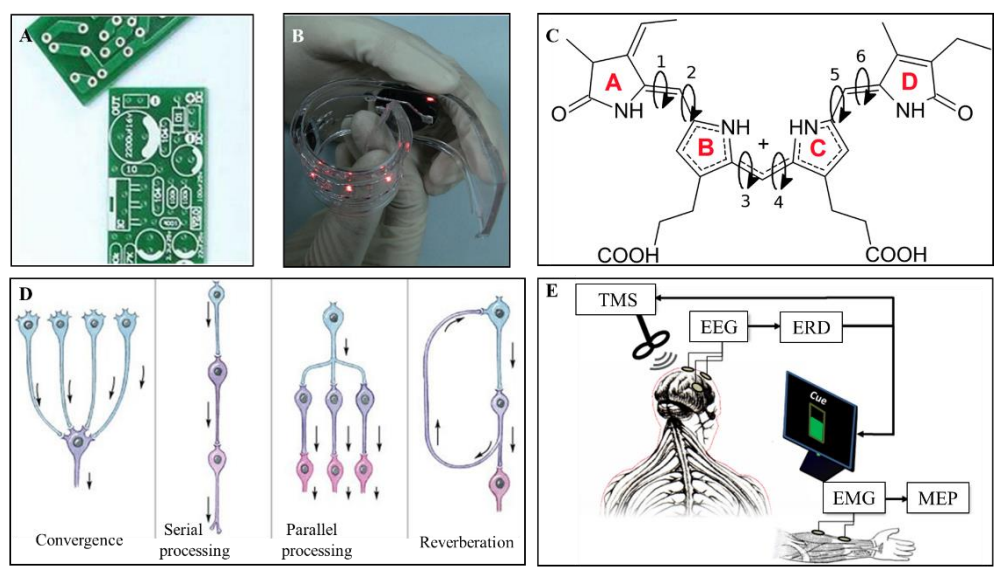

**Fig 1**. Respective kinds of electrical circuits. (A) Traditional rigid PCB circuits (http://cn.bing.com); (B) Soft electronics [14]; (C) Molecular switch (http://cn.bing.com); (D) Several typical types of neural circuits (http://cn.bing.com); (E) Brain-computer interface [11]. Note: TMS: transcranial magnetic stimulation; ERD: event-related desynchronization; MEP: motor evoked potential; EEG: electroencephalogram; EMG: electromyogram. (Note: Figures reproduced with permission).

Liquid metal refers to alloys or metals with a low melting point, which can maintain liquid phase around room temperature [15]. It has recently been introduced into soft electronics and biomedical field [15]. Liquid metal sensors [16-17], memristors [18], diodes [19] and electrodes [20-22] have already been proposed for health-monitoring and disease treatment. Recently, some researches have been devoted to the study of liquid metal droplets, which show great potential in self-powered devices [23, 24] and phagocytosis [25]. Yang et al introduced millimeter-scale LMDs as thermal switches, unlocking new possible solutions for thermal management [26]. Tang et al electrically controlled the size and rate of LMDs formation [27]. Others fabricated non-stick LMDs by coating polytetrafluoroethylene particles [28] or graphene [29] on NaOH-treated LMDs. Chen et al. found that graphene-coated LMDs can be used as droplet-based floating electrodes [29]. Sivan et al. coated LMDs with n-type and p-type semiconducting nanopowders to study their electronic properties and electrochemical properties [30].

Different from previous studies where all LMDs were contacted to realize electrical conduction, droplet circuit means that the circuit is composed of discrete LMDs and can operate well just as traditional circuits do. For this purpose, we would speculate that when the gap between two LMDs is small enough, electrical signal can be transported even though they are not connected. This assumption is dependent on the theory of quantum tunneling effect which is a physical phenomenon that a micro-particle such as an electron can tunnel through a barrier that it classically could not surmount. Thus, droplet circuits can transfer electrical signal even though LMDs are separated in space. Zhao et al had ever put forward a transformable soft quantum device based on liquid metal [31]. They introduced that liquid metal droplets can be adopted to create tunnel junction and defined four configurations of all-soft quantum devices, as shown in Fig. 2(E). Based on their research, this article further explores the potential of liquid metal droplets in constructing droplet circuits and preliminarily demonstrates its probability. Compared to the traditional rigid or soft circuits, liquid metal droplet circuits provide more flexibility without the problem of circuit break.

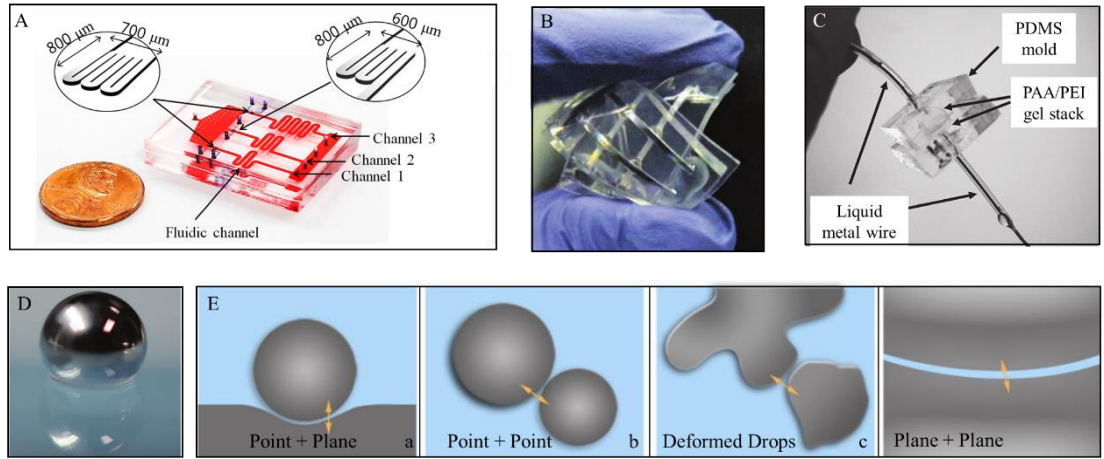

**Fig 2**. (A) Liquid metal pressure sensor [16]; (B) Liquid metal memristor [18] (Reproduced with permission); (C) Liquid metal diode [19] (Reproduced with permission); (D) Liquid metal droplet [28] (Reproduced with permission); (E) Four configurations of all-soft liquid metal quantum devices based on tunneling effect [31] (Reproduced with permission).

## 2. Electrical Conduction via Quantum Tunneling Effect

Quantum tunneling phenomenon cannot be explained by classical mechanics, occurring only at quantum scale. Compared to classical mechanics, matter in quantum mechanics owns the properties of waves and particles, involving the Heisenberg uncertainty principle [32-34]. Heisenberg uncertainty principle is that one can never exactly know the position and speed of a particle at the same time. This can be described by the following inequality:

$$\sigma_x \sigma_p \geq \frac{h}{4\pi} \tag{1}$$

Here, $\sigma_x$ is the standard deviation of position, $\sigma_p$ refers to the standard deviation of momentum, and $h$ is Planck constant.

For the inequality to be satisfied, no probability could be exactly zero. Thus, events which seem to be impossible in classical mechanics become possible in quantum mechanics. This can be used to explain quantum tunneling phenomenon. For instance, Fig. 3(A) shows a rectangle barrier. Even though the energy of a particle is lower than the threshold energy of the barrier, it is possible that the particle appears at the other side of the barrier. That feels like there is a tunnel at the bottom of the barrier. The energy of the tunneled particle is the same as before but the amplitude of possibility is declined.

Quantum tunneling is essential in many occasions, including nuclear fusion in sun [35], astrochemistry in interstellar clouds, quantum biology, tunnel diodes (Fig. 3(B)) [36], tunnel junctions (Fig. 3(C)) [37], scanning tunneling microscope (Fig. 3(D)) and quantum computing etc. Tunnel junction is that two conductors are separated by a thin insulator to create a simple barrier between them. It can be applied to measure voltages and magnetic fields.

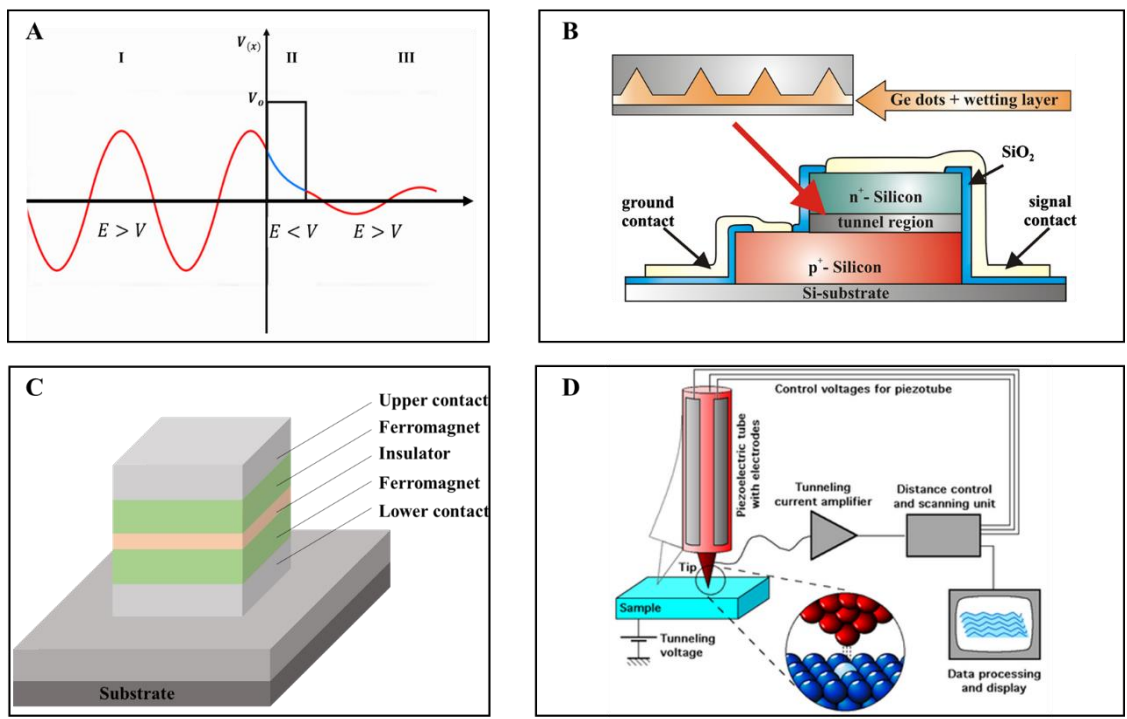

**Fig 3.** Illustration of quantum tunneling effect and conventional typical application devices. (A) Quantum tunneling effect; (B) A Ge quantum dots interband tunneling diode [36]; (C) Diagram of a ferromagnet tunnel junctions; (D) Scanning tunneling microscope (http://cn.bing.com). (Note: Figures reproduced with permission).

## 3. Fabrication of LMDs and Composing of Droplet Circuits

So far, ways for the fabrication of LMDs have already been relatively diverse and easily available, including methods like sonication, molding, and flow-focusing [27]. In order to prevent the formation of oxide layers on the surface of LMDs, water films and nano particles are applied [28, 38]. The regular and stable distribution of LMDs can be induced by external magnetic or electrical field. Recently, Yu et al. created a method named as suspension 3D printing [39] which can help quickly realize various three dimensional droplet patterns. They successfully patterned LMDs into self-healing hydrogel (Fig. 4(A)) and studied the relationship between the process parameters, supporting gel concentration, and the deposited micro-droplet geometry. Before that, the present lab also successfully prepared LMDs in large scale through a channelless fabrication method [40], as shown in Fig. 4(B). And Tian et al proposed a microfluidic chip for liquid metal droplets generation and sorting. Their system could manipulate these neutral liquid metal droplets in nonconductive fluid [41]. All those studies provide viable methods for the fabrication of LMDs and construction of droplet circuits and more efforts should be made to improve the technology.

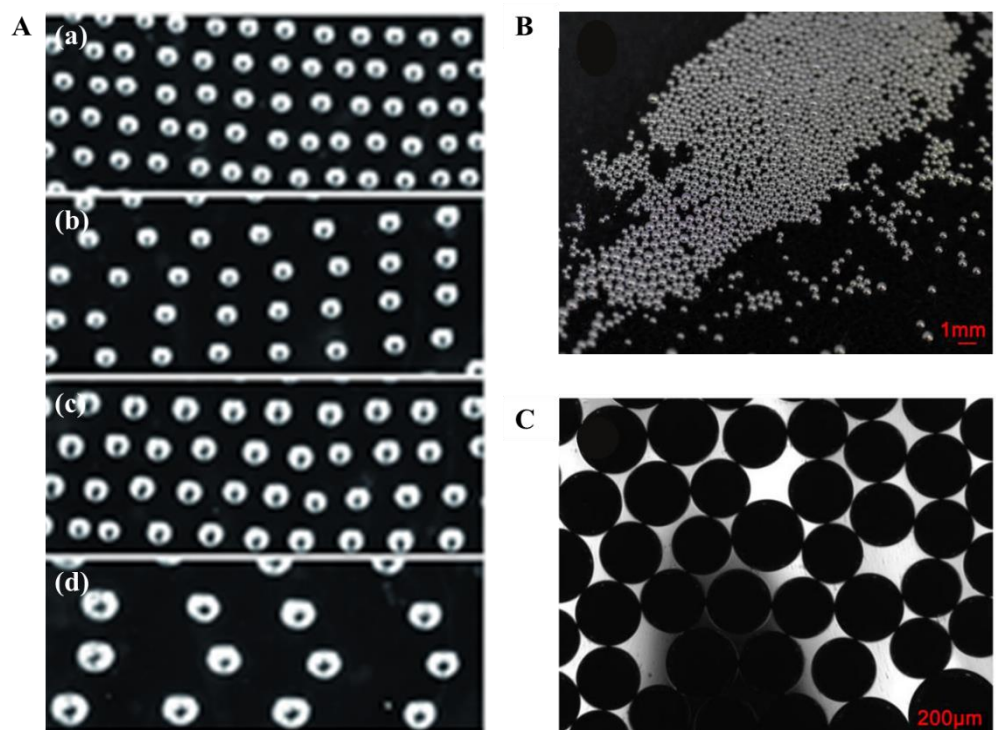

**Fig 4**. (A) LMDs floating in hydrogel with varied diameters; (a) 60um, (b) 90um, (c) 160um, (d) 210um [39]; (B) LMDs assembly fabricated through injection way [40]; (C) Microscopic image showing a single layer of liquid metal micro-droplets closely sitting together with thin interface between each other [40]. (Note: Figures reproduced with permission).

**4. Droplet Circuits in Solution**

**4.1 Mechanism of liquid metal based droplet circuits**

Droplet circuits are mainly composed of discontinuous LMDs, ions and conductive wires such as carbon nanotubes and operate in electrolyte solution. LMDs and carbon nanotubes are mixed up and cooperate to make the circuit electrically connected. Carbon nanotubes have unique electrical, thermal and chemical properties, showing great potential in nanoelectronics [42], especially for transistor application owing to benign carrier mobility and velocity. Therefore, carbon nanotubes can be selected for connecting droplet circuits.

Fig. 5(A) shows the electrical conduction of droplet circuits. LMDs in droplet circuits are surrounded by carbon nanotubes and ions and electrons transfer along the pattern of LMDs with the assistance of carbon nanotubes and ions. There are two ways for LMDs to communicate in electrolyte solution, as shown in Fig. 5(B). When two LMDs are sitting close enough, quantum tunneling effect will happen and electrons can transfer from one to another possibly. If the distance of two LMDs becomes a little farther than tunneling effect can support, carbon nanotubes floating between them serve as transfer stations for electrons to flow. Through quantum tunneling effect and carbon-nanotube transfer stations, those discrete LMDs are electrically connected and the whole circuit can work well. In general, LMDs are randomly arranged in electrolyte solution. Through applying voltage to those LMDs, they will be organized in order and make the circuit connected (Fig. 5(C)).

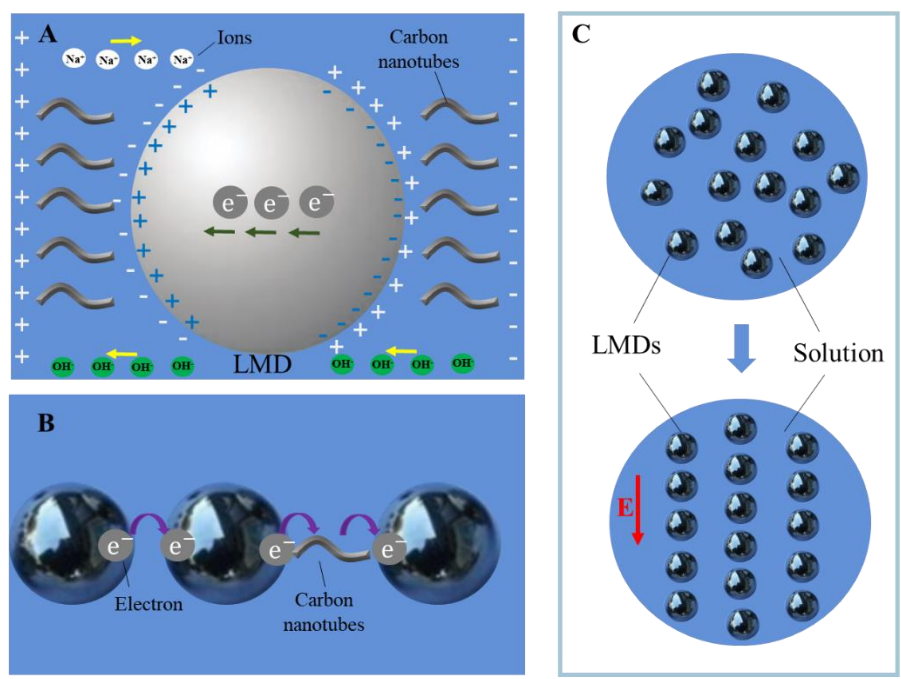

**Fig 5**. (A) Electronic conduction in droplet circuits composed of LMDs, ions and conductive carbon nanotubes; Tunneling effect between two LMDs; (B) Combination of two communicating ways for LMDs in a circuit; (C) LMDs arranged with and without voltage.

As the resistance of LMDs is smaller than other regions in electrolyte solution, the current will transfer mainly along LMDs regularly. In addition, one can change the pattern of LMDs to control the conducting direction of current. That is, given specific topological or directional design, the liquid metal droplets composing the circuit can achieve desired or regulative functions of electron transport through conductive routes or quantum tunneling effect. Some of the newly emerging needs such as room temperature computing or brain-like chip can possibly enabled based on such kind of unconventional electrical circuits.

**4.2 Configuration of liquid metal droplet circuit**

*4.2.1 Low-dimensional droplet circuits*

Zero-dimension (0D) droplet circuit refers to that only one LMD works. Fig. 6(A) shows schematic of 0D circuit. The LMD is immersed in NaOH solution and replaces part of the wire. When the switch is closed, the lamp can be lit up.

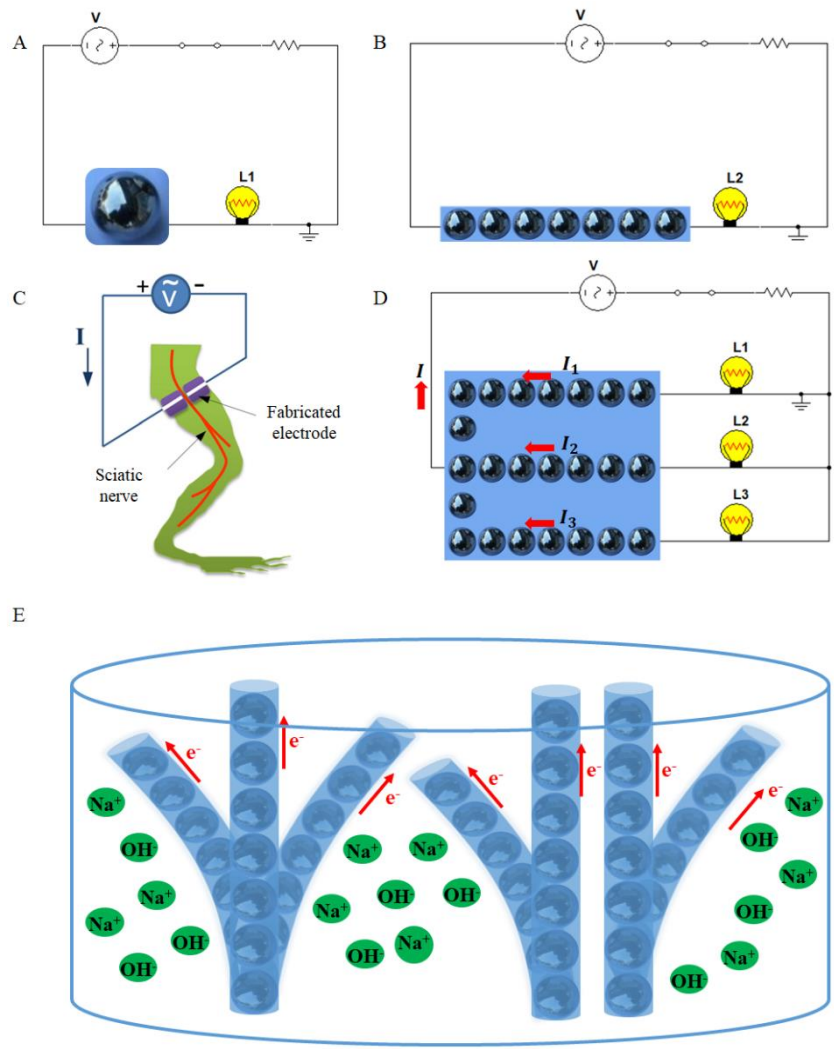

**Fig 6**. (A) 0 D droplet circuit; (B) 1 D droplet circuit; (C) Loading the electrical stimulations to the sciatic nerve of a frog through liquid metal [43] (Reproduced with permission); (D) 2 D droplet circuit; (E) Topological structure of droplet circuit.

Increasing the number of LMDs and arranging them in a line, then a one-dimension (1D) droplet circuit can be formed. The line of LMDs can be inserted into a traditional circuit as an alternative to the original wire, as shown in Fig. 6(B). Similarly, one can arrange LMDs in a plane with specific pattern to realize a two-dimension (2D) droplet circuit (Fig. 6(D)). In principle, this circuit should also satisfy Kirchhoff's Current Law:

$$\sum I_{in} = \sum I_{out} \tag{2}$$

Here, $I_{in}$ is the current flowing into a node and $I_{out}$ is the current flowing out from the node.

Kirchhoff's Law states that charge input at a node is equal to the charge output. Therefore, currents of the circuit in Fig. 6(D) should meet the following equation:

$$I = I_1 + I_2 + I_3 \tag{3}$$

Due to existence of the electrolyte in the solution, charge loss is inevitable. But it is small enough to be neglected.

Furthermore, LMDs can be arranged in special topological structure to fabricate three-dimension (3D) even dynamically transformable droplet circuit, as shown in Fig. 6(E). These 3D droplet circuits are flexible and able to transport electrons along the desired direction. Topological droplet circuits perfectly imitate the connection between neurons, which is beneficial to artificial neural connection.

The combination of liquid metal and biological neuro circuit has been demonstrated before by Zhang et al [43]. As shown in Fig. 6(C), they injected liquid metal in the left and right side of the sciatic nerve near the femur of an interceptive lower part of a bullfrog body. It was found that the two electrodes successfully conducted the electrical-stimuli signals to the nerves, which proves the feasibility of applying liquid metal to electrical circuits.

*4.2.2 High-dimensional droplet circuits*

The further development of droplet circuits may be more than three dimensions. In this regard, LMDs can be encapsulated into elastic tubes together with carbon nanotubes and electrolyte solution together to replace conventional rigid metal wires, as shown in Fig. 7(D). Moreover, since the pattern of LMDs can be changed by external magnetic or electrical field, the circuit will become more fantastic to dynamically change over the time. Applying electrical or magnetic field to LMD circuits, the structure of the circuits can be changed as required. For instance, LMDs can be induced to move by magnetic field when coating them with ferromagnetic materials (Fig. 7(A), (B)) [44]. With the help of aluminum, one can realize reliable motion control of the liquid metal droplets in electrical field [45]. Fig. 7(C) presents the sequential movement of a liquid metal droplet propelled by external electrical field. The velocity of Al/EGaIn and Ni/Al/EGaIn droplets in NaOH solution under different voltages has been measured and calculated, as shown in Fig. 7(E) [46]. It is apparent that the LMDs speed up with the increase of the voltage. Moreover, the current of droplet circuits composed of LMDs in NaOH solution is relatively stable under constant voltages (Fig. 7(F)), demonstrating promising application of droplet circuits [47].

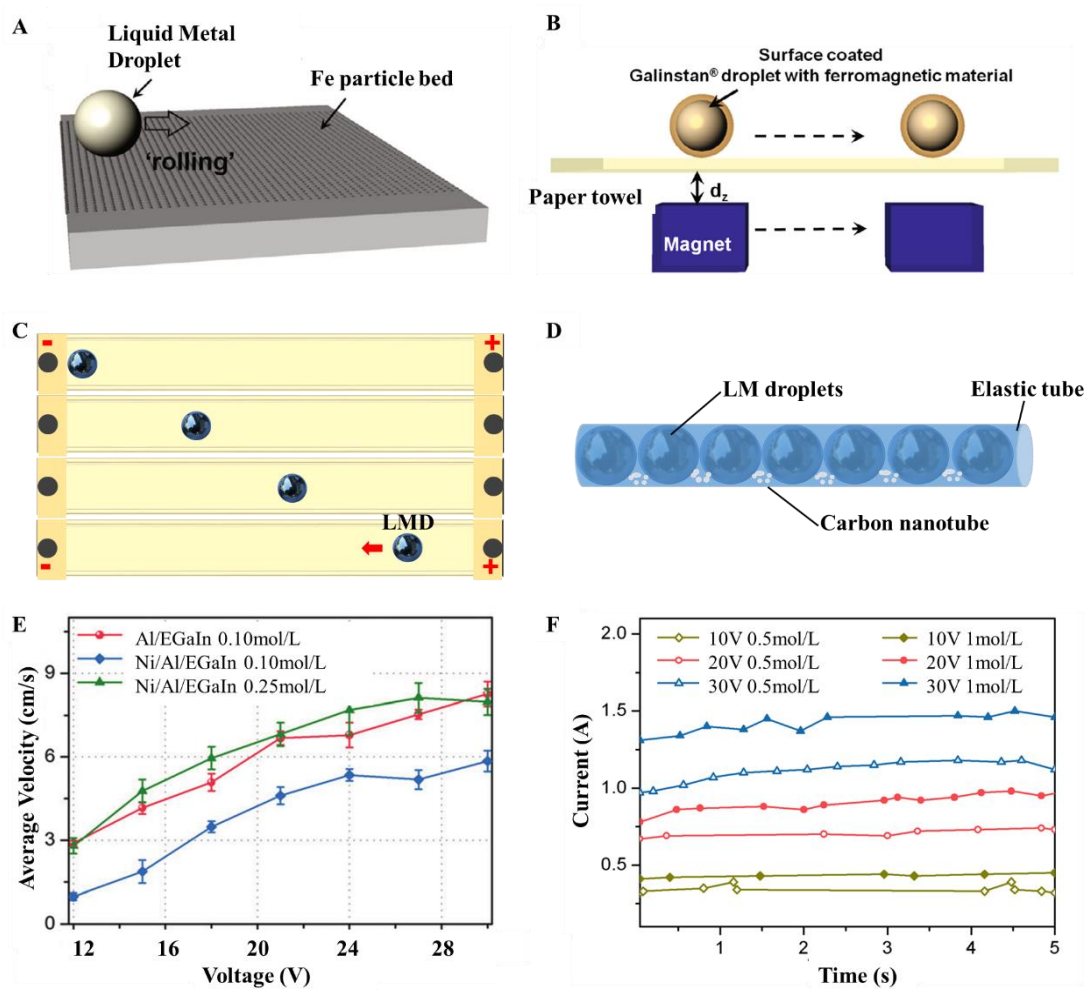

**Fig 7**. (A) Fabrication of Fe-coated LMDs [44] (Reproduced with permission); (B) Surface coated LMD with ferromagnetic materials induced by magnetic field [44] (Reproduced with permission); (C) Sequential photos of a LMD's motion in electrical field; (D) Liquid metal droplet wire. (E) Average velocity of LMDs in a straight channel containing NaOH solution under exposure to different voltages, and the concentrations belong to NaOH solution [46] (Reproduced with permission); (F) The electric current in response to different voltages, and the concentrations belong to NaOH solution [47] (Reproduced with permission).

**5. Discussion**

This article exploits the basic features of the droplet circuits and evaluates the probability of liquid metal droplets enabled functional circuits. Obviously, droplet circuits display unique advantages which may not easily be offered by conventional strategies. First of all, droplet circuits show highly parallel electrical transport capability. In addition, such circuits own excellent flexibility, self-healing capability and stretchability. Different from existing liquid metal soft electronics, LMD circuits consist of spatially transformable discrete liquid metal droplets. It can tolerate greater elastic deformation and be shaped into various electrical patterns.

Further, droplet circuits are fault-tolerant and self-error-correcting. Since LMDs can transport electrons while they are disconnected from each other, droplet circuit may not face errors of circuit break. Even though the distance between LMDs can be changed by external factors such as strain and twist, carbon nanotubes will keep the circuit electrically connected. It is apparent that droplet circuit appears more robotic than traditional circuit, which is very much like the working style of a biological neuro circuit.

Particularly, droplet circuits could run in the wet environment of electrolyte solution. Electrical conduction involves not only LMDs and carbon nanotubes, but also ionic conduction. Ions in electrolyte solution play the role of electron transportation along with carbon nanotubes when the quantum tunneling effect of LMDs does not work.

Last but not least, the fabrication of liquid metal based droplet circuit is overall not complicated since LMDs are self-assembly in the environment of magnetic or electrical field. The circuit configurations are in dynamic change at the same time.

From practical aspect, due to its unique advantages such as high flexibility and benign electrical conductivity, the liquid metal based droplet circuits can be possibly applied to make artificial retina or cochlea as an alternative to conventional rigid wires and electrodes. Nerve repair and neural connection is another promising development for such droplet circuits. Efforts had ever been made to explore the possibility of liquid metal neural restoration [43]. Unlike former trial, the current droplet circuit offers more electron transport channels such as: electrically conductive liquid metal, ionic conduction and nano wires etc. Therefore, it would better serve as the medical needs. Another worth of mentioning potential of the droplet circuit lies in its role in constructing computing chip or devices which are different from the classical framework. With discrete LMDs, the droplet circuit offers opportunities as quantum processor or to carry out quantum calculating which may help design future quantum computer.

## 6. Conclusion

This work presents the new conceptual droplet circuit which is enabled from the liquid metal and demonstrates its potential roles in electronics field. The advantage of droplet circuits lies in that it well addresses the problem of circuit break and allows more flexibility and self-healing feature, showing great potential in developing smart soft electronics which could well imitate the biological neuro circuits in nature. In addition, droplet circuits are potential for possible application in the coming quantum calculation. Such droplet circuits are easily fabricated and self-error-correcting. This shows its promising value in molding large scale application technologies. Further research could be conducted to fabricate stable and nano-scale liquid metal droplets in electrolyte solution and test the electrical properties of those droplets. Clearly, science, technology and application along this direction are waiting for systematic investigation and integration together in the coming time.

**Acknowledgment**

This work is partially supported by the NSFC Key Project under Grant No.91748206, Dean's Research Funding of the Chinese Academy of Sciences and the Frontier Project of the Chinese Academy of Sciences.